\documentclass[useAMS,usenatbib]{mn2e}

\usepackage{amssymb}
\usepackage{amsfonts}
\usepackage{mncite}
\usepackage{epsfig}
\usepackage{psfig}

\begin{document}

\title[Dust filtration by disc-embedded planets]
{Dust filtration at gap edges: Implications for the spectral energy distributions of discs with embedded planets}

\author[W.K.M. Rice, Philip J. Armitage, Kenneth Wood \& G. Lodato]{W.K.M. Rice$^1$\thanks{E-mail: wkmr@roe.ac.uk}, Philip J. Armitage$^{2,3}$, Kenneth Wood$^4$ and G. Lodato$^5$\\
$^1$ SUPA\thanks{Scottish Universities Physics Alliance},
Institute for Astronomy, University of Edinburgh, Blackford Hill, Edinburgh, EH9 3H \\
$^2$ JILA, Campus Box 440, University of Colorado, Boulder CO 80309 \\
$^3$ Department of Astrophysical and Planetary Sciences, University of Colorado, Boulder CO 80309 \\
$^4$ SUPA$\dagger$, School of Physics and Astronomy, University of St Andrews, North Haugh,
St Andrews, Fife KY16 9SS \\
$^5$ Institute of Astronomy, Madingley Road, Cambridge, CB30HA}

\maketitle

\begin{abstract}
The spectral energy distributions (SEDs) of some T Tauri
stars display a deficit of near-IR flux that could be a
consequence of an embedded Jupiter-mass planet partially
clearing an inner hole in the circumstellar disc. Here, we
use two-dimensional numerical simulations of the planet-disc
interaction, in concert with simple models for the dust
dynamics, to quantify how a planet influences the dust
at different radii within the disc. We show that pressure
gradients at the outer edge of the gap cleared by the planet
act as a filter - letting particles smaller than a critical
size through to the inner disc while holding back larger particles
in the outer disc. The critical particle size depends upon the
disc properties, but is typically of the order of 10 microns.
This filtration process will lead to discontinuous grain populations
across the planet's orbital radius, with small grains in the
inner disc and an outer population of larger grains. We show
that this type of dust population is qualitatively consistent
with SED modelling of systems that have optically thin inner 
holes in their circumstellar discs. This process can also
produce a very large gas-to-dust ratio in the inner disc, potentially
explaining those systems with optically thin inner cavities that
still have relatively high accretion rates.
\end{abstract}


\section{Introduction}
A fraction of T Tauri stars, including TW~Hya \citep{calvet02,uchida04}, 
GM~Aur \citep{marsh92,koerner93,rice03}, CoKu Tau/4 \citep{forrest04,quillen04, dalessio05} and 
DM~Tau \citep{calvet05}, present spectral energy distributions (SEDs) that combine little 
excess flux (above the photospheric level) at near-IR wavelengths with 
robust excesses at $\lambda \gtrsim 10 \mu{\rm m}$. A similar effect
is seen in at least one Herbig Ae/Be star, HD100546
\citep{bouwman03,grady05}. Since circumstellar 
discs are generally most optically thick in the inner, hottest regions, 
this observation is noteworthy and indicative of some process that 
can create an optically thin inner cavity within the disc. One 
exciting possibility is that the cavity arises from the tidal 
truncation of the inner disc by a massive planet \citep{goldreich80,lin86}, 
in which case the SED observations provide unique constraints on the 
time scale and frequency of giant planet formation. However, non-planetary 
explanations are also possible. Dust can grow rapidly to form solid 
bodies that contribute negligible opacity in any observed waveband, while 
larger bodies may themselves suffer destructive collisions or rapid orbital 
decay due to aerodynamic drag against the gaseous disc \citep{weidenschilling77}.
Photoevaporation of the outer disc by ionizing stellar
radiation can also starve the inner disc of gas, yielding a
short inner hole phase even in the absence of a planet
\citep{alexander06}. 
Protoplanetary disc models that include some or all of these 
effects \citep{takeuchi01,kenyon04,dullemond05,takeuchi05} admit the 
formation of dusty rings and inner holes, even in the absence of planets 
and gas. As improved SEDs and spectra from {\em Spitzer} become available, 
more detailed theoretical study of both classes of model (pure disc 
and disc plus embedded planet) is warranted in order to determine whether 
these models can be distinguished without direct imaging data.

A sufficiently massive planet (or brown dwarf) can probably create 
an inner cavity in which both the gas and dust surface densities 
are extremely low. However, of the four aforementioned 
T~Tauri stars -- around which inner disc `holes' 
are inferred from the lack of near-IR emission -- only CoKu Tau/4 
appears to be entirely devoid of circumstellar material within 1~AU 
of the star. In the case of TW~Hya there is both ongoing low-level 
gas accretion \citep{muzerolle00}, and evidence (from the existence of 
a silicate emission feature at 10$\mu$m) for a sub-lunar mass of micron 
sized dust at small radii \citep{calvet02}. Both 
GM~Aur and DM~Tau are Classical T Tauri stars, with 
estimated accretion rates on to the stellar photosphere variously 
estimated at $\dot{M} \approx 10^{-9} \ M_\odot {\rm yr}^{-1}$ to 
$10^{-8} \ M_\odot {\rm yr}^{-1}$ \citep{gullbring98,hartmann98,white01}.
Unfortunately \citet{najita03} failed to detect CO emission 
from an inner disc around GM~Aur, so there are no observational 
clues as to the {\em gas} surface density that must accompany this 
accretion. However, for these accretion rates protoplanetary disc models 
predict gas surface densities at $r \approx 0.1 \ {\rm AU}$ of 
200-2000~gcm$^{-2}$ \citep{bell97}. Although disc models are subject 
to large uncertainties arising from the unknown efficiency of angular 
momentum transport within protoplanetary discs, the robust accretion 
signatures seen for these stars imply that an inner gas disc must be 
present. 

Within a planetary interpretation of the observed SEDs, the presence 
of inner gas discs and ongoing stellar accretion is not in itself 
surprising. Numerical simulations \citep{artymowicz96,lubow99} show 
that the tidal barrier erected by a massive planet is quite leaky. 
For Jupiter mass planets gas can readily overflow the gap, and 
while much of this gas is accreted by the planet in excess of 
10\% of the accretion rate at large radius can replenish the 
inner disc and ultimately accrete on to the star \citep{lubow06}. 
This is consistent with GM~Aur being a Classical T~Tauri star, 
but the lack of dust opacity then requires that the planet needs 
to be relatively more effective at holding back dust -- i.e. it needs 
to filter the gas as it passes the tidal barrier. In this paper, we show 
that the required filtration may occur as a consequence of the 
sharp pressure gradient that exists near the outer edge of the gap. 
The pressure gradient leads to super-Keplerian rotation of the gas, 
which in turn allows standard aerodynamic coupling to solid bodies \citep{weidenschilling77,paardekooper06} 
to inject angular momentum and hold back the inward flow of all but the 
smallest solid particles. We estimate the efficiency of this process, and argue 
that it allows relatively low mass planets -- which are observed 
\citep{marcy05} to be much more common than those planets with 
$M_p > M_J$ that would start to cut off accretion entirely -- to 
nevertheless create dust-poor inner cavities. Within this 
scenario, the variation in the observed properties of T~Tauri 
stars with weak near-IR emission could reflect different mass 
embedded planets -- relatively low in the case of GM~Aur and 
DM~Tau, and substantially higher in the case of CoKu Tau/4.

\section{Gas-dust dynamics}
\label{gasdust}
The disc gas and the embedded dust particles interact via a drag force 
\citep{whipple72, weidenschilling77} that is a consequence of the
different gas and dust orbital velocities.  The Keplerian velocity, $V_K$,
at a radius, r, around a star of mass $M_*$ is 
\begin{equation}
V^2_K = \frac{G M_*}{r}.
\label{kepler}
\end{equation}
If $P$ is the gas pressure and if $\rho$ is the gas density, the azimuthal component 
of the gas velocity, $v_\phi$, in centrifugal equilibrium 
differs from the Keplerian velocity due to the influence of the gas
pressure gradient and is given by
\begin{equation}
\frac{v^2_\phi}{r} = \frac{V^2_K}{r} + \frac{1}{\rho}\frac{dP}{dr}.
\label{vphi}
\end{equation}
The dust particles, on the other hand, are not influenced by pressure forces
and, in centrifugal equilibrium, orbit at the Keplerian velocity, $V_K$.

The effect of the drag force depends on the size of the dust particles
relative to the mean free path of the gas molecules.  If we assume that
the gas is primarily molcecular hydrogen, the mean free path, $\lambda$, is 
\begin{equation}
\lambda = \frac{m_{H_2}}{\rho A} \approx \frac{4 \times 10^{-9}}{\rho} {\rm cm}
\label{mfp}
\end{equation}
where $m_{H_2}$ is the mass of the hydrogen molecule and $A$ is its cross-section ($A = \pi a^2_o \approx 7 \times 10^{-16} {\rm cm}^2$).  The drag force, 
$F_{\mathrm{D}}$ is then 
\begin{equation}
{\mathbf F}_{\mathrm{D}}=-\frac{1}{2}C_{\mathrm{D}}\pi
a^2\rho u^2\hat{\mathbf u},
\label{FD}
\end{equation}
where ${\mathbf u}$ is the relative velocity between the gas and dust particles,
$u=|{\mathbf u}|$, $\hat{\mathbf u}={\mathbf u}/u$, $a$ is the
mean radius of the dust grains and $C_{\mathrm{D}}$ is the drag
coefficient, given by
\begin{equation}
C_{\mathrm{D}} = \left\{ \begin{array}{ccc}
        \displaystyle \frac{8}{3}\frac{c_s}{u} & ~~~~~~~~ & a<9\lambda/4\\
\\
        24R_e^{-1}                              & ~~~~~~~~ & R_e<1 \\
\\
	24R_e^{-0.6}                            & ~~~~~~~~ & 1<R_e<800\\
\\
	0.44                                    & ~~~~~~~~ & R_e > 800.
	 	\end{array} \right.
\label{CD}
\end{equation}
In equation (\ref{CD}), $R_e$ is the Reynolds number which can be shown to
be given by (see for example \citet{rice04})
\begin{equation}
R_e=4\left(\frac{a}{\lambda}\right)\left(\frac{u}{c_s}\right).
\label{Reynolds}
\end{equation}

\begin{figure}
\centerline{\psfig{figure=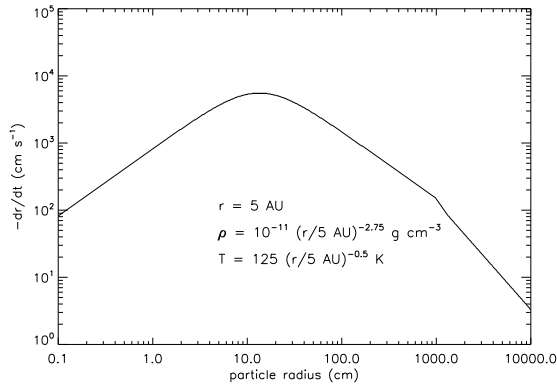, width=75mm}}
\caption{Radial drift velocity against planetesimal size at 5 AU in a disc
with a density and temperature, at 5 AU, of $10^{-11}$ g cm$^{-3}$ and $125$ K 
respectively.}
\label{radin}
\end{figure}

In general, the gas pressure in protoplanetary discs decreases with increasing
radius.  The pressure gradient is therefore negative and the
gas velocity is sub-Keplerian. The drag force then acts to slow the
dust particles down causing them to spiral in towards the central star.
One can then determine the radial drift velocity of dust particles by self-consistently
solving the radial and azimuthal components of the momentum equation as
outlined in \citet{weidenschilling77}. Figure \ref{radin} shows
an example of the inward radial velocity (-dr/dt) against particle radius 
at $5$ AU in disc with a density of $\rho(r) = 10^{-11} (r/5 \mathrm{AU})^{-2.75}$ g cm$^{-3}$ 
and a temperature of $T(r) = 125 (r/5 \mathrm{AU})^{-0.5}$ K.
The radial velocity is clearly strongly dependent on the particle size, and
can reach values of $\sim 10^4$ cm s$^{-1}$.

The inward migration of dust particles, however, only occurs if the the
pressure in the disc decreases monotonically with increasing radius. If
there are any regions of the disc where the gas pressure
increases with increasing radius, the gas velocity will become super-Keplerian,
and the drag force will cause dust particles
to move outwards towards the pressure maxima.  This could occur in the
presence of self-gravitating spiral structures \citep{haghighipour03, rice04},
in the presence of vortices \citep{klahr03}, or if an embedded planet has
opened a gap in the disc \citep{paardekooper04,paardekooper06}. At the outer edge
of a gap opened by a planet, 
the drag force will cause dust particles to migrate up the
gap edge, away from the star. If, however, there is still gas accretion
through the gap, which is expected for Jupiter mass companions 
around solar-like stars, then dust particles moving too slowly 
up the gap edge will be dragged by the gas into the gap. The gap edge
will therefore act like a filter, allowing only particles of a certain
size through into the inner disc.  Even if the inner gas disc has not
drained onto the central star, there could still be a substantial
difference between the dust populations in the inner and outer disc
and this could have a noticeable effect on the disc SED.

To investigate this further, we need an approximation for the pressure
gradient at the outer edge of a gap opened by an embedded planet.  We therefore
perform two-dimensional simulations of gaseous discs with embedded planets
to determine the approximate disc structure.

\section{Planet-disc simulations}
We simulate a planet embedded in a gaseous disc using the Zeus code 
\citep{stone92}. We use two-dimensional coordinates ($r$,$\phi$) and a 
resolution of $n_r = 400$, $n_\phi = 400$.  The computational domain
extends, in code units, from $r_{\rm in} = 0.5$ to $r_{\rm out} = 5$ and we impose ouflow
boundary conditions at both $r_{\rm in}$ and $r_{\rm out}$. 

We assume that the disc is isothermal with a radially dependent sound 
speed given by $c_s(r) = 0.05 r^{-1/2}$. We adopt units in which $G = 1$
and the sum of the mass of the central star and embedded planet is $M_* +
M_{\rm pl} = 1$.  Since the disc thickness $h/r$ is related to the sound
speed through $h/r \approx c_s / V_K$, we have a disc
thickness of $0.05$ at $r = 1$.  We model angular momentum transport
using a kinematic viscosity, $\nu$, that operates only on the azimuthal
component of the momentum equation \citep{papaloizou86}.  The initial
disc surface density, $\Sigma$, is taken to have a radial dependence
of $\Sigma(r) \propto r^{-1}$. Since $\dot{M} \propto \nu \Sigma(r)$
and since we would expect $\dot{M}$ to be constant we assume
that, $\nu(r) \propto r$.  We normalise the viscosity using the 
standard alpha formalism, $\nu = \alpha c_s h$ \citep{shakura73}, and
assume that at $r = 1$, $\alpha = 10^{-3}$.  

\begin{figure}
\centerline{\psfig{figure=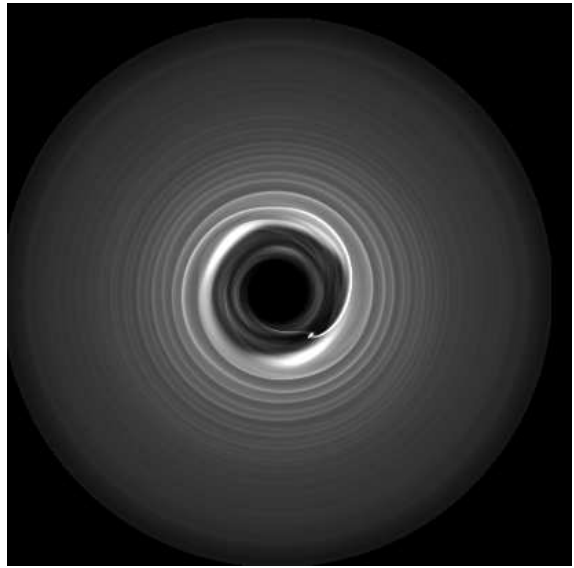,width=75mm}}
\caption{Disc surface density structure for the simulation in which the
embedded planet has a mass of $M_{\rm pl} = 10^{-3}$.  It is clear that the 
interaction between the planet and the disc gas has significantly depleted
the surface density in the inner disc.}
\label{plangap}
\end{figure}

\begin{figure}
\centerline{\psfig{figure=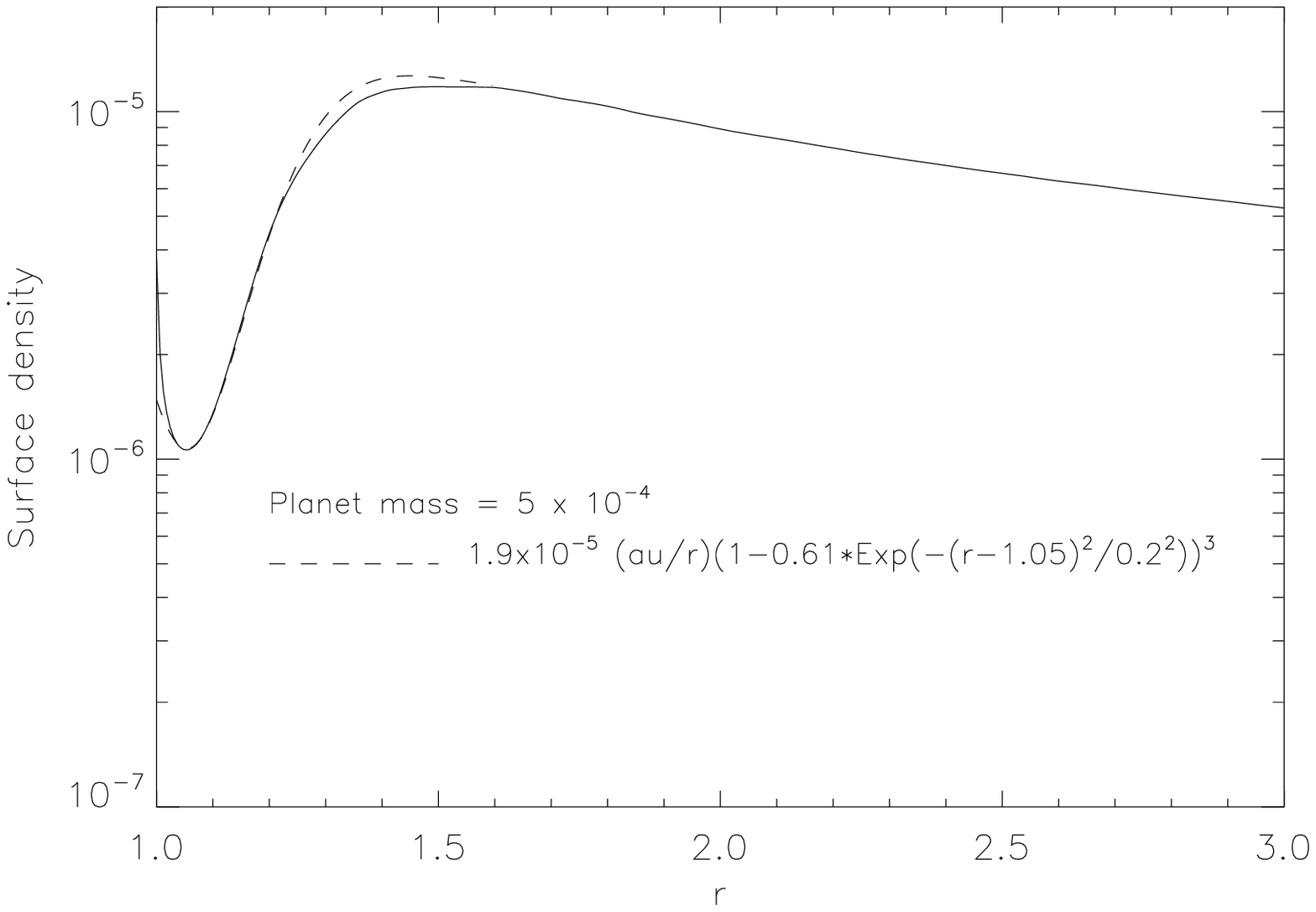, width=75mm}}
\centerline{\psfig{figure=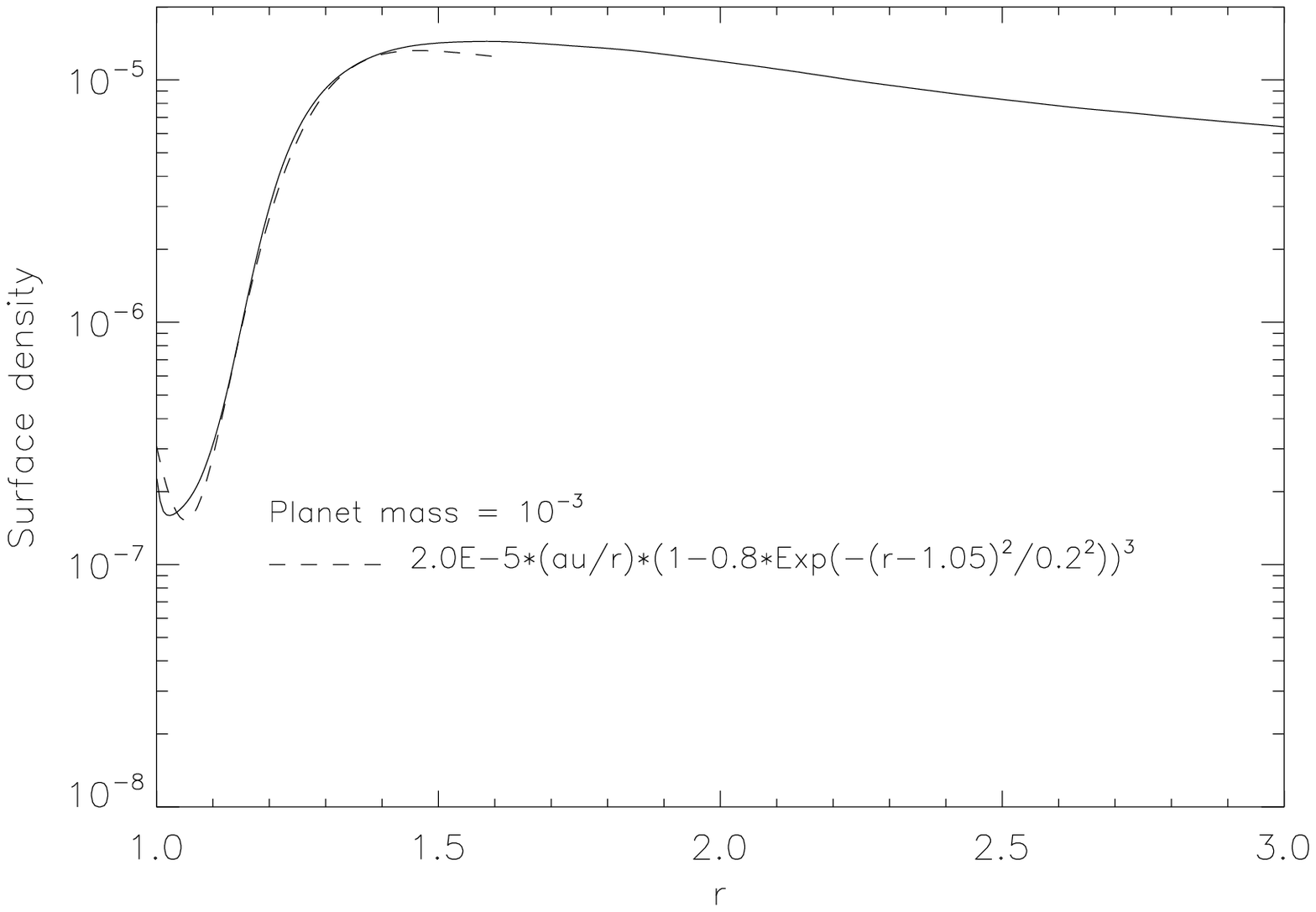,width=75mm}}
\centerline{\psfig{figure=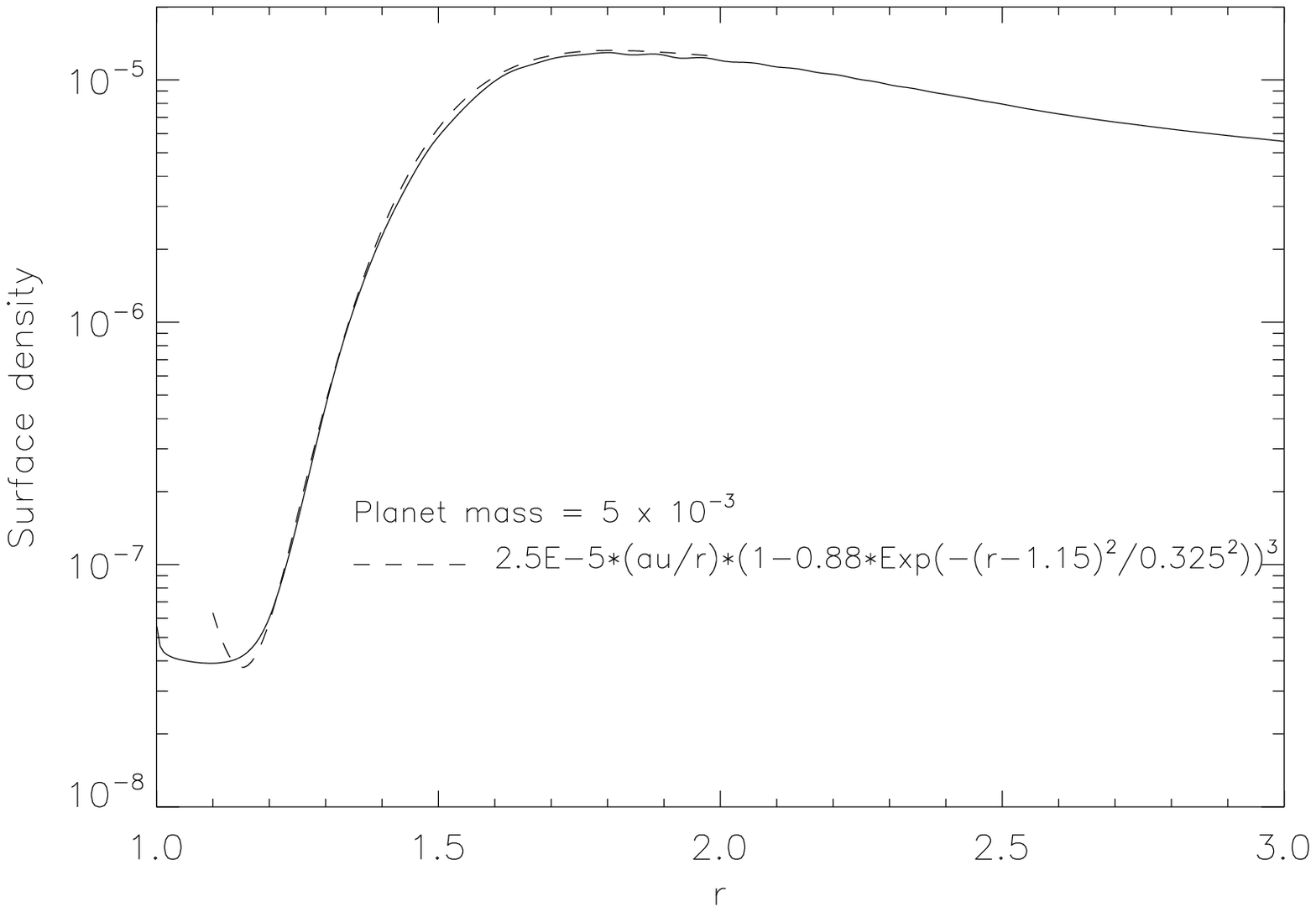,width=75mm}}
\caption{Disc surface density profiles for the three planet masses 
considered in the gasdynamic simulations.  
As expected, the gap depth increases as the planet mass increases. The dotted line
also shows an approximate analytic fit to the profile of the gap edge which can
be used to determine how the presence of a planet influences the interaction between the disc
gas and the embedded solid particles.}
\label{surfdens}
\end{figure}

The planet is located at $r = 1$ and we consider planet masses of $M_{\rm pl} = 5 \times 10^{-4}$, 
$M_{\rm pl} = 10^{-3}$, and $M_{\rm pl} = 5 \times 10^{-3}$.  
The simulations are evolved for $\sim 1000$ planetary
orbital periods. Figure \ref{plangap} shows the surface density structure of the simulation with 
$M_{\rm pl} = 10^{-3}$ and illustrates that the inner regions of the disc have been significantly depleted.  
Figure \ref{surfdens} shows the surface density profiles for the
three different planet masses.  Also shown (dashed line) is an emperical fit to
the gap edge. The surface density is normalised such that the total disc mass
within $r=100$ would be $10^{-2}$ in code units.  
It appears that the chosen function is a reasonable fit
to the gap edge, especially where the surface density is steepest, which is the region that
would have the largest pressure gradient and hence would be the region where the drag
force would have the largest influence on the embedded dust particles. From Figure 
\ref{surfdens} a general form for the function representing the gap edge would be
\begin{equation}
\Sigma_{\mathrm edge} = \Sigma_o r^{-1}\left\{ 1 - f {\mathrm \exp} 
\left[ \frac{-(r - r_{\rm gap})^2}{d_{\rm gap}^2} \right] \right\}^3,
\label{gaussian}
\end{equation}
where $r_{\rm gap}$ is the inner radius of the gap edge, and $d_{\rm gap}$ is the gap scalelength.
From the fits in Figure \ref{surfdens}, the gap scalelength is somewhat greater for 
$M_{\rm pl} = 5 \times 10^{-3}$ 
than for the lower planet masses, but we assume in general that 
$d_{\rm gap} = r_{\rm gap}/5$. The gap depth itself
depends on the value of the coefficient $f$. A large $f$ value results in a deeper gap with a wider edge, and
would correspond to a more massive planet. If we use Figure \ref{surfdens} as a guide, 
$f \sim 0.6$ would correspond to a planet mass of $M_{\rm pl} = 5
\times 10^{-4}$, while $f \sim 0.9$ would correspond to a planet mass
of $M_{\rm pl} = 5 \times 10^{-3}$.  These simulations have all been
performed in scale free code units. To convert to real units, we
simply need to introduce length and mass scales. For the problem we
are considering here we will generally assume a mass scale of $1
M_\odot$ giving a total disc mass of $10^{-2} M_\odot$ and 
planet masses of about  $0.5$, $1$, and $5$ Jupiter
masses.  The length scale we introduce depends on the assumed semimajor axis of
the planet. In the rest of this paper we will assume a length scale
of $4$ au, resulting in a semimajor axis of $4$ au
and, from Figure \ref{surfdens}, an outer gap edge at $\sim 6$ au. 

\section{Dust migration at gap edges}
To determine how gas drag influences dust particles at the edge of a gap opened by an embedded
planet, we need to know the gas pressure gradient. Equation (\ref{gaussian}) gives an 
approximate form for the surface density structure at the edge of a gap.  The volume density is
then, $\rho = \Sigma / h$ where $h = c_s r/ V_K$.  The sound speed, $c_s = \sqrt{k T/ \mu m_H}$ 
where $\mu = 2.4$ is the mean molecular weight, $k$ is Boltzmann's constant
and the temperature, $T = 280 (AU/r)^{1/2} {M_*/M_\odot}$ \citep{hayashi81} with $M_* = 1 M_\odot$. 
The pressure is then $P = \rho c_s^2$ and the pressure gradient at the gap edge
can be determined exactly since we have a fully analytic form for the 
pressure as a function of radius.  We should, however, note that since the
discs we consider here have optically thin
inner regions, the outer edge of the gap will intercept more stellar
flux and will be hotter than if the inner disc were optically thick
\citep{dullemond04,akeson05}. A higher temperature will produce a larger pressure gradient. 
This temperature increase will, however, decay very quickly within the gap
edge, and so may not be significant at the location where the pressure
gradient is largest. Even if the
temperature is slightly enhanced, a steeper temperature gradient will
act to reduce the effect of the increased temperature. Overall we
would not expect the effect of the increased temperature at the edge
of the gap to be particularly significant. Once the pressure gradient is
calculated, the technique described in \S \ref{gasdust}
can then be used to determine the radial velocity of solid particles at any point on the 
edge of the gap. 

\begin{figure}
\centerline{\psfig{figure=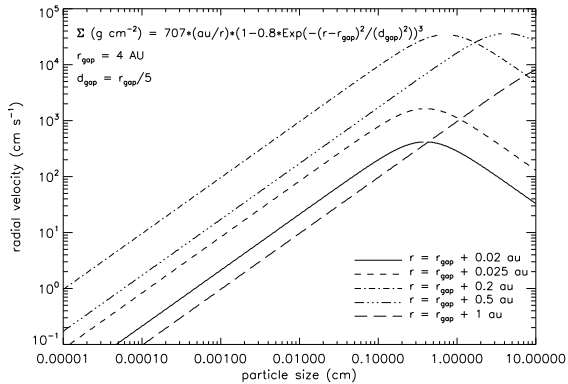,width=75mm}}
\caption{Radial velocity against particle size at various locations on
the edge of a gap in
which the gap edge is described by equation (\ref{gaussian}) with $f =
0.8$.  As $r$ increases, the radial velocities initially increase
reaching a maximum when $r = r_{\rm gap} + 0.2 au$. As $r$ increases
further, the radial velocity decreases for particles smaller than
the particle with the largest radial velocity ($a < 1 - 10$ cm). 
For particles small enough to contribute to the SED ($a < 0.1$ cm),
there is therefore a location where they have a maximum outward radial
velocity.}
\label{radvelf8}
\end{figure}

The radial velocity of the solid particles will vary considerably along the edge of the gap. 
This is illustrated in Figure \ref{radvelf8} which shows the radial velocity against
particle size at the edge of a gap in which, $r_{\rm gap} = 4$ AU,
$d_{\rm gap} = 0.8$ AU, and $f = 0.8$.  In this case the radial
velocity is positive, as opposed to Figure \ref{radin} in which the
radial velocity is negative. The surface density in Figure
\ref{radvelf8} is also normalised such that in the absence 
of a gap the total disc mass within $100$ AU would be $0.05 M_{\odot}$.  The different curves in
Figure \ref{radvelf8} correspond to different locations (i.e., differents values
of $r$) on the gap edge. All the curves in Figure
\ref{radvelf8} have the same form as in Figure \ref{radin}. At a given location, the 
radial velocity intially increases
as the particle size increases until it reaches a peak radial velocity, after which the radial
velocity decreases with increasing particle size.  As $r$ increases, the curves initially move
upwards and the radial velocity increases for all particle sizes (solid line, small dash line, and
dash-dot line) reaching a maximum when $r = r_{\rm gap} + 0.2$ AU.  As
$r$ increases beyond this point (dash-dot-dot-dot line, and long dash line), the curves move
to the right (i.e., the particle size at which the radial velocity is maximal increases), and
the maximum radial velocity decreases. 
The radial velocity for all particles smaller than the 
particle with the largest radial velocity therefore decreases as $r$ increases. Beyond the peak of
the gap edge, the gradient changes sign, the radial velocity changes sign, and solid particles
will drift inwards, rather than outwards. 

\begin{figure}
\centerline{\psfig{figure=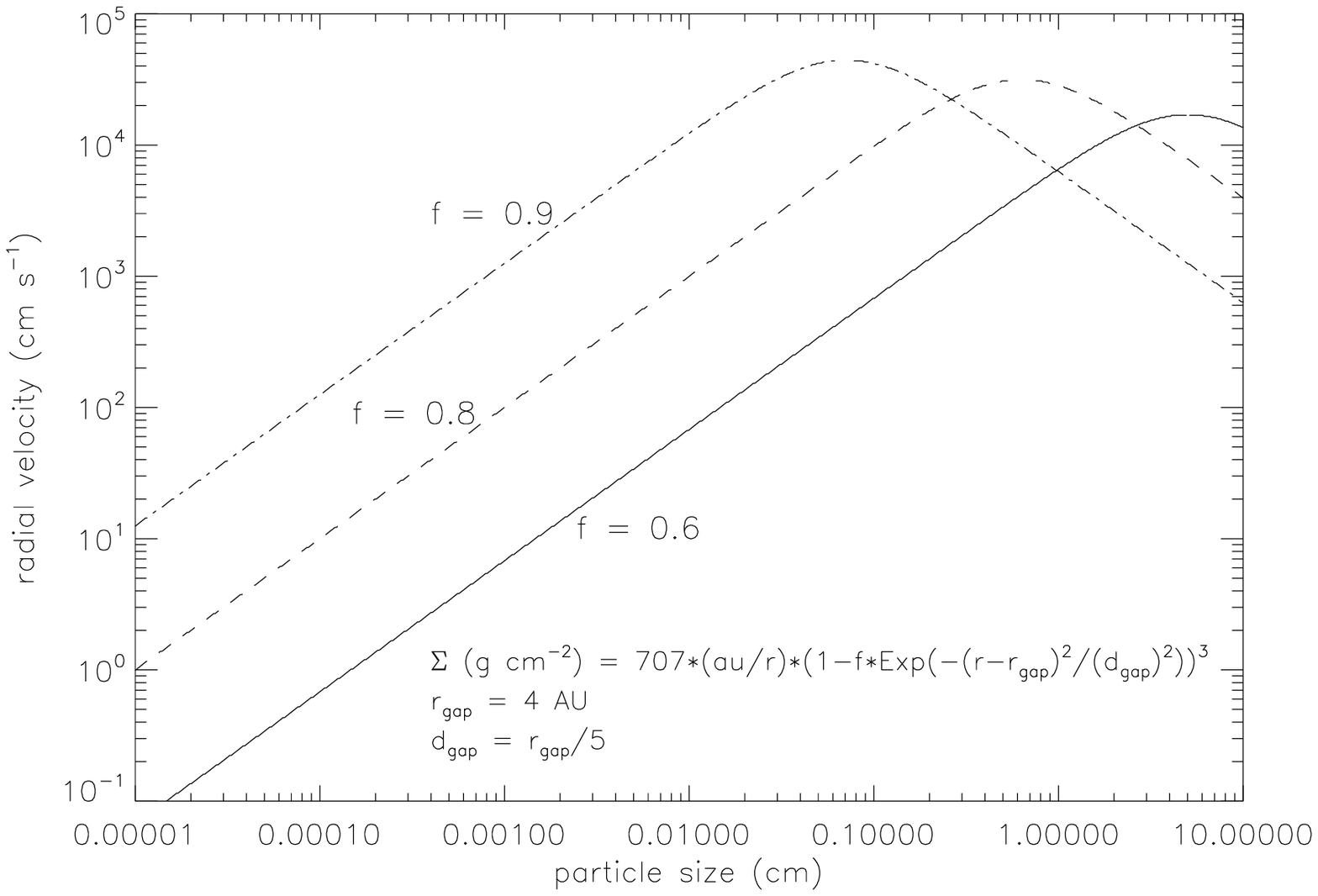,width=75mm}}
\centerline{\psfig{figure=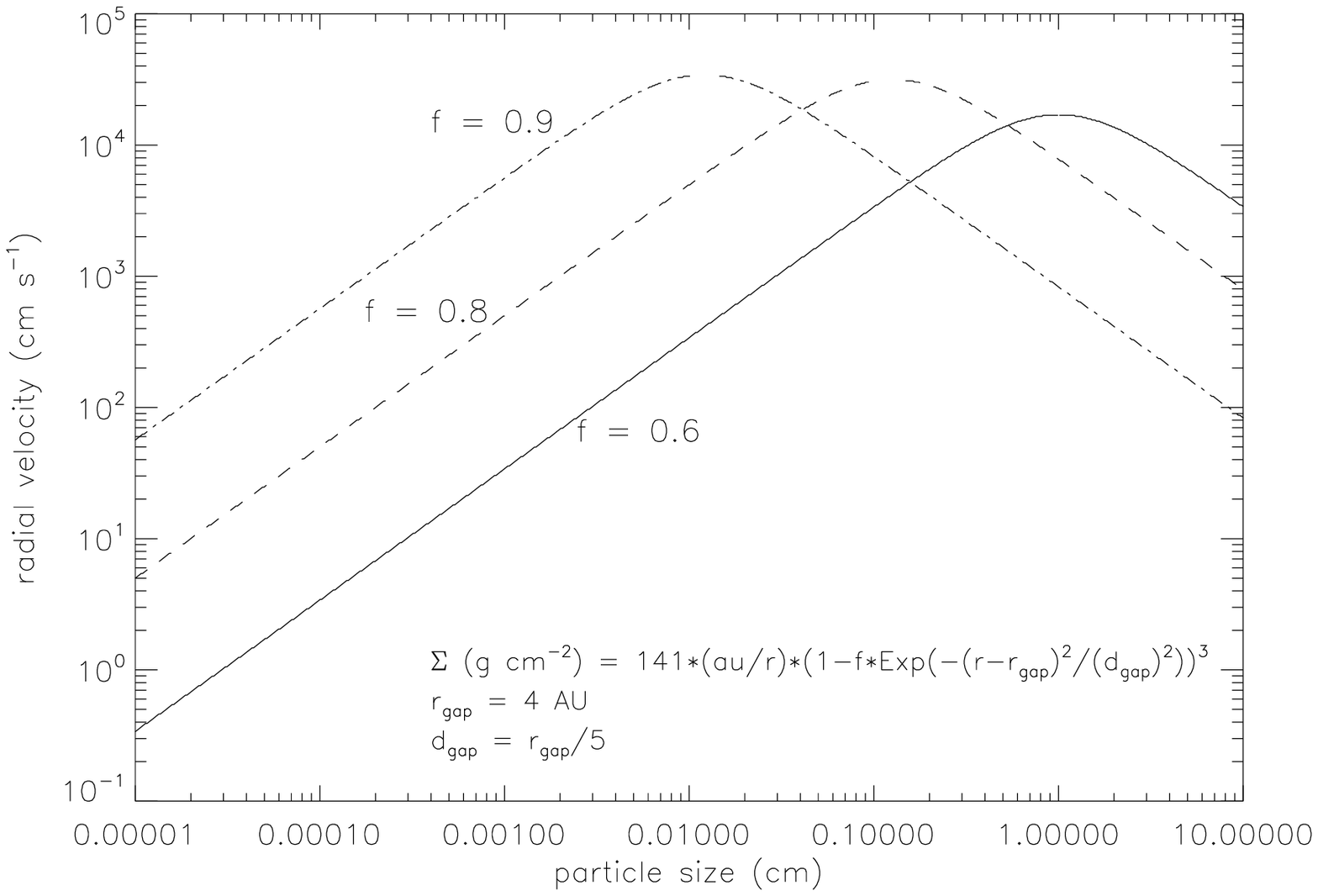,width=75mm}}
\caption{Radial velocity against particle size at the edge of a gap opened by an
embedded planet.  The two panels correspond to surface densities that differ by a factor
of 5.  The three curves in each panel correspond to different assumed gap depths, the deeper
gaps assumed to result from the presence of a more massive planet.  In both panels the peak
radial velocity moves to smaller particles sizes as the gap depth increases resulting
in an increase in the radial velocity of all particles smaller than the peak particle size. 
The gap edge will therefore act to filter out the larger particles
that have radial velocities larger than the inward radial velocity of the gas, allowing only
the smaller particles into the gap.}
\label{radveldrho}
\end{figure}

What we wish to determine is the maximum outward radial velocity for particles that would
contribute to the SED (eg., $a < \sim 0.1$ cm). This can be determined by simply varying $r$ until
these particles achieve their maximum velocity. In Figure \ref{radvelf8} this 
would correspond to the dash-dot line ($r = r_{\rm gap} + 0.2$ AU). The two panels in
Figure \ref{radveldrho} show the maximum outward radial velocity as a function of particle size
for two different surface densities and for three different $f$ values.  
In the upper panel the total disc mass within 
$100$ AU would be $0.05 M_\odot$, while in the lower panel the total mass would be
$0.01 M_\odot$.  In both figures $r_{\rm gap} = 4$ AU and $d_{\rm gap} = 0.8$ AU. 
The three curves in each figure correspond to $f = 0.6$, $f = 0.8$, and $f = 0.9$. 
Based on the simulation results shown in Figure \ref{surfdens}, these could correspond
to planet masses of $0.5$ M$_{\rm Jup}$, $1$ M$_{\rm Jup}$, and $5$ M$_{\rm Jup}$. 

The curves in Figure \ref{radveldrho} show that as $f$, or planet mass, increases, the peak radial
velocity moves to smaller particle sizes.  All particles smaller than the peak particle
size then have increased radial velocities.  The radial velocities in Figure
\ref{radveldrho} are, however, measured in the frame in which the gas radial velocity is zero.
If there is no accretion (i.e. the radial gas velocity is zero relative to the central star), 
all particles would be drifting up the gap edge, away from the central star.   If there was
some gas accretion, then particles with radial velocities smaller than the radial velocity
of the gas will be dragged into the gap. 

It is somewhat hard to quantify the radial
gas velocity. The presence of a planet can significantly influence the 
velocity structure of the gas in the vicinity of the planet
\citep{kley99, bate03}.  We are, however, interested in the radial
velocity of the gas as it accretes through the outer edge of the gap.  
Simulations \citep{kley99, bate03} show that in general the planet
does not influence this material significantly, at least for planets
with masses below a few Jupiter masses \citep{lubow06}.  The material that is most
influenced by the planet is material that has already accreted into
the gap. Most of this material also ends up within the planet's hill
sphere, ultimately accreting onto the planet. The material that we are
considering here is material that accretes through the gap into the
inner disc, and it appears that this is largely unaffected by the
planet, at least in the vicinity of the gap edge.  The gas radial
velocity through the edge of the gap will therefore be
determined by the disc viscosity, and is related to the mass accretion rate 
through $\dot{M} = -2 \pi R \Sigma v_r$.

For an accretion rate of $10^{-9} \ M_\odot {\rm yr}^{-1}$ the radial 
gas velocity in the absence of a gap would be $\sim 1$ cm s$^{-1}$ for the
higher of the surface densities (upper panel) 
in Figure \ref{radveldrho} and $5$ cm s$^{-1}$ 
for the lower of the surface densities (lower panel) 
in Figure \ref{radveldrho}. If the mass transfer through
the gap edge is unaffected by the planet, the radial gas velocity should then increase as
the surface density decreases. When $f = 0.6$, the surface density when the radial
velocity of the particles maximises is $\sim 7$ times lower than at the peak of the gap edge.  The radial
gas velocity should then be $\sim 7$ times higher, which for the upper panel
of Figure \ref{radveldrho}, and for the mass transfer rate assumed above, would give 
$v_r \sim 7$ cm s$^{-1}$.  
This means that only particles smaller than $\sim 10$ microns will
be swept into the gap by the radial motion of the gas since these particles have outward radial
velocities less than $\sim 7$ cm s$^{-1}$.  For $f = 0.9$, the surface
density changes by a factor of $860$ and, if the same applies, then the radial gas
velocity would then be $\sim 860$ cm s$^{-1}$.  The maximum particle size that can populate the
inner disc is then $\sim 5$ microns, comparable to that obtained when $f = 0.6$. 
An equivalent result is obtained if we consider the lower 
panel in Figure \ref{radveldrho}.  The surface density being 5 times smaller than in
the upper panel results in the particles' radial velocities 
also increasing by a factor of $5$. For a
given mass accretion rate, however, the radial gas velocity also increases by a factor of $5$ and hence
the size of particles that would be swept into the gap is roughly indepedent of the assumed
surface density.  

On the other hand, numerical simulations by \citet{lubow99} suggest that the mass accretion rate
through a gap actually decreases with increasing planet mass, at least for planets
with masses in excess of $1$ Jupiter mass.  In this case the
gravitational torques from the planet act to inhibit flow through the
gap \citep{lubow06}. The rate of decrease with mass appears
to be exponential, with the mass accretion rate decreasing by almost an order of magnitude
as the planet mass increases from $1$ to $6$ Jupiter masses.  If this is indeed
the case, then the size of particles that will be swept into the gap by gas accretion
should decrease with increasing planet mass.  Again considering the upper panel of 
Figure \ref{radveldrho}, if the three curves correspond to planet masses of $\sim 0.5 M_{\rm Jup}$,
$\sim 1 M_{\rm Jup}$, and $\sim 5 M_{\rm Jup}$ then we may expect the 
gas radial velocity to decrease by an order of magnitude
as $f$ varies from $0.6$ to $0.9$. The maximum particle
size swept into the gap would then decrease by an order of magnitude. For the gas radial
velocities considered above, we might expect the maximum particle size to decrease from
$\sim 10$ microns to a few tenths of a micron as $f$ increases from $0.6$ to $0.9$ corresponding
to the planet mass increasing from $0.5$ M$_{\rm Jup}$ to $5$ M$_{\rm Jup}$. 

Since the above process results in only the smallest dust grains being swept into the gap, 
we would expect the dust to gas mass ratio to be different in the inner disc, 
compared to the outer disc.  It is, however, difficult to make an independent measurement
of the gas and dust masses, and hence determining the different dust to gas mass ratios is
probably not possible.  The presence of a silicate feature in protostellar disc spectral
energy distributions (SEDs) does, however, depend on the grain size distribution.
The presence of large grains tends to wash out the silicate feature. As we discuss
in the next section, SED modelling of those systems with near infrared deficits suggest that
the grain size distribution in the inner disc differs from that in the outer disc in a way
that is consistent with dust filtration at a gap edge.

\section{Discussion}
Although the suggestion that the observed near-IR deficits in some TTauri SEDs
could result from an embedded Jupiter-like planet clearing the inner disc is 
exciting, there is, as yet, no clear evidence that this is indeed the case. Other
non-planetary possibilities, for example rapid grain growth, could
also result in optically thin regions that may produce SEDs largely consistent
with that observed. 

A sufficiently massive planet does, however, have a real effect on the structure of
the gaseous disc which may influence the distribution of solid particles. This
effect could, in turn, influence the nature of the resulting SED.  A $\sim$ Jupiter
mass planet should open a gap in the disc, the inner region of which could clear to form
an inner hole.  The outer edge of this gap (or hole) will have a pressure that increases
with increasing radius, increasing the net outward force on the gas and resulting in
super-Keplerian gas velocities.  The resulting drag force on the solid particles then
results in these particles drifting up the gap edge, away from the central star. If, however,
there is still some gas accretion into the gap - which is expected for $\sim$ Jupiter-mass planets \citep{lubow99} -  
solid particles with radial velocities smaller
than the radial gas velocity will populate the inner region of the disc.  For disc properties appropriate
for TTauri stars, the radial velocity increases with particle size for particles smaller
than $\sim 1$ cm. This means that only the smallest particles will have outward radial
velocities small enough to be dragged into the gap by the accreting gas.

Gas accretion through the gap edge therefore acts to filter solid particles, with only the smallest
particles making it into the gap.  The exact size range that is influenced by the gas depends
on the properties of the gas disc and on the radial gas
velocity. Assuming a functional form for the gap
edge (motivated by two-dimensional, hydrodynamic simulations of embedded Jupiter-like planets) we 
show that if the inward gas velocity is $\sim 1 - 10$ cm s$^{-1}$, only $\sim$ micron and smaller grains
will be able to populate the inner disc.  If the gas accretion rate is modified by sufficiently massive planets,
as suggested by simulations \citep{lubow99}, the size of particles that may populate the inner disc
should decrease with increasing planet mass. Systems, such as CoKu Tau/4, that appear to have almost no
material in the inner disc may indicate the presence of a massive companion while those, such as GM Aur
and TW Hya, that appear to still have some material in the inner disc, may have a companion with a mass
that allows some material to continually replenish the inner disc.

What is attractive about this possibility is that SED modelling suggests that system with near-IR deficits
do indeed have two populations of dust particles.  This spatial segregation of dust within discs has been
inferred in the TW Hya disc \citep{calvet02}, and radiation transfer models of new Spitzer Space Telescope
IRS spectra of the CoKu Tau 4, GM Aur and DM Tau discs also require inner holes with small grains in the
inner disc, and a larger grain population in the outer disc \citep{dalessio05, calvet05}.  The silicate 
feature present in the IRS spectra provide the clues that the dust grains in the inner disc are smaller
than in the outer disc.  As more data comes in from Spitzer, the number of disc systems showing evidence for inner
holes is increasing \citep{lada06,muzerolle06}. To date, most candidates have IRAC and MIPS photometry,
so only the presence of inner disc clearing can be inferred. Follow up observations with IRS will enable us
to determine whether the spatial segregation of dust sizes is a common feature and thereby further test
our dynamical models for disc-planet interactions and dust filtration.

The SED modelling of systems like GM Aur and DM Tau not only require a
population of small grains in the inner disc, but also require that the
mass, or surface density, is also low enough for the inner disc to be
optically thin. Accretion onto a planet can reduce the amount
of mass reaching the inner disc by up to $90 \%$ \citep{lubow06}, but
this is probably insufficient to produce an optically thin inner
disc. Typically the optical depth will need to be reduced by a factor
of $\sim 10^4$, a factor of $1000$ more than can be produced by the planet
alone.  If solid particles are filtered at the gap edge, 
the amount of mass reaching the inner disc will depend on the
size distribution of the solid particles.  Interstellar medium (ISM) grains are
generally assumed to have sizes between $0.005 \micron$ and $1
\micron$ and a size distribution of $n(r) \propto
r^{-3.5}$ \citep{mathis77}. Grain growth can however change this
distribution quite significantly. If there is constant replenishment
of the ISM grains, this can steepen to $n(r) \propto r^{-4}$
\citep{mizuno88}, in which case most of the mass will be in small
grains. If, however, there is no source of ISM grains, the final size
distribution can be much flatter, approaching $n(r) \propto
r^{-2}$. A relatively shallow size distribution (e.g., $ \propto
r^{-3}$) will already have a reduced opacity, independent of the
presence of a planet, but, as shown below (see also Figure
\ref{SEDgaps}), a further reduction is needed to match the
observed SEDs.  Such a reduction can be obtained through the filtration process described here.

\begin{figure}
\centerline{\psfig{figure=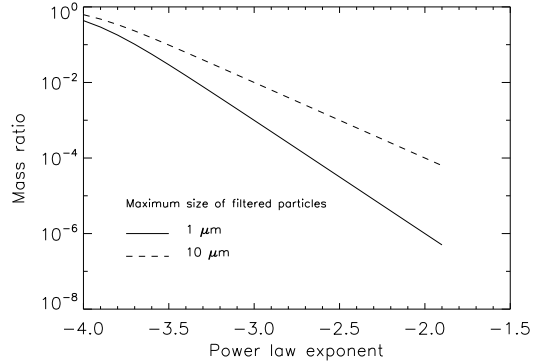,width=75mm}}
\caption{Ratio of the mass of particles with sizes below $1 \micron$ (solid line) and 
$10 \micron$ (dashed line) to the total mass of solid particles, plotted
  against the exponent of the power-law size distribution.  If the
  maximum size of particles filtered at the gap edge is $1 \micron$,
  and if the power-law exponent is $-3$, the mass of material
  accreting into the gap would be reduced by a factor of $10^3$.  If
  the planet accretes $90 \%$ of this mass, the amount of material
  reaching the inner disc would be reduced by $10^4$, sufficient to
  produce an optically thin inner disc.}
\label{massratfilt}
\end{figure}

\begin{figure}
\centerline{\psfig{figure=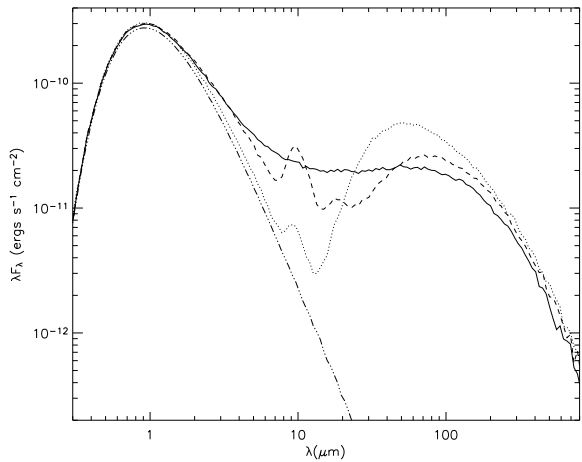,width=65mm,angle=90}}
\caption{SED models that illustrate how filtration of dust particles at the outer
edge of a gap can influence the resulting SED.  These models are computed assuming
a 300 au flared disc of total mass $0.01 M_\odot$ that is passively heated by a star
with $T_{\rm star} = 4000$ K, and $R_{\rm stars} = R_{\odot}$.  The solid curve
is the SED resulting from a dust model with a grain size distribution of $n(r) \propto
r^{-3}$ with minimum and maximum grain sizes of $0.005 \micron$ and $1$ mm respectively.
The other two models have the same grain population in the outer disc, beyond $5$ au, but have ISM-like grains
in the inner disc with depletion factors of $10^4$ (dashed line) and $10^6$ (dotted line).
Notice the near-IR deficit and prominent silicate features at $10 \micron$ and $20 \micron$ for
the SED models with ISM-like grains and inner disc depletions, consistent with observations
of a number of T Tauri systems.}
\label{SEDgaps}
\end{figure}

The amount of material that can reach the inner disc will depend on
the maximum size that can accrete through the gap edge, and on the
size distribution of the particles in the outer disc.  Figure
\ref{massratfilt} shows the ratio of the mass of particles smaller
than $1 \micron$ (solid line) and $10 \micron$ (dashed line) to the
total mass of solid particles, plotted against the exponent of the
power-law size distribution.  In this figure we assume a minimum size
of $0.005 \micron$ and a maximum size of $1$ mm.  For steep size 
distributions most of the mass is in small grains, and even if the 
maximum size of filtered particles is $1 \micron$, the inner disc 
would remain optically thick.  It is also likely that in this case, grain
coagulation would soon produce
large grains in the inner disc \citep{dullemond05}. From SED
modelling, however, we know that systems like GM Aur, TW Hya and DM
Tau  must have a size distribution that extend to larger sizes in the
outer disc and,
since these systems are reasonably evolved, there is unlikely to be a
significant new source of ISM grains.  We might, therefore, expect the size
distribution in the outer disc to be flatter than the ISM grain size
distribution \citep{mizuno88}. It has already been shown that in some
systems, flatter grain size distribution (e.g., $n(r) \propto
r^{-2.5}$) provide better fits to the SEDs than ISM size 
distributions \citep{dalessio01}.  Figure \ref{massratfilt} shows that
a size distribution of $n(r)
\propto r^{-3}$ and a maximum filtered size of $1 \micron$, would
reduce the mass of solid material accreting through the gap edge by a 
factor of $10^3$.  If the planet accretes $90 \%$ of this material, 
the amount of solid material reaching the inner disc would be reduced by a
factor of $10^4$, sufficient to produce an optically thin inner
disc. 

This process would also produce a very different dust-to-gas
ratio in the inner disc, compared with the outer disc. That systems 
like GM Aur and DM Tau have optically thin inner discs but are still
accreting at  $\dot{M} \ge 10^{-9} \ M_\odot {\rm yr}^{-1}$
suggests that there is still a substantial amount of gas in the inner
disc. The optically thin inner disc requires a reduction in solid
particles of order $10^4$, while the ongoing accretion requires a far
smaller reduction in the amount of gas in the inner disc. This is
at least qualitatively consistent with the suggestion that the solid
particles are filtered at the outer edge of the gap which, together
with the subsequent accretion onto the planet reduces the mass of
solid particles by a factor of $\sim 10^4$, while at least
$10 \%$ of the gas is allowed through into the inner disc \citep{lubow06}.
Such a reduction in the solid particle surface density in the inner
disc also means that subsequent grain coagulation is probably not very
important. \citet{dullemond05}, using a relatively simple one-particle
model \citep{safronov69}, show that at $\sim 1$ au
grain coagulation can produce $\sim 500 \micron$ particles within a few
hundred years for reasonable surface densities.  If the mass that
can be swept up by a growing particle is then reduced by a factor of
$10^4$, one might assume that the maximum size that this particle could achieve is then reduced
by a factor of $20$ (i.e., a maximum size of $\sim 25$ microns).  This, however, is a
significant overestimate since the amount of mass swept up by a
growing particle depends strongly on the size of the particle, or its
collision cross-section. In the simple one-particle model, there is
almost no grain growth if the surface density is reduced by a factor
of $10^4$ from an initial density that would be appropriate for the inner
regions of a TTauri disc.  Although this is a fairly simple model, it at least
suggests that if the amount of mass reaching the inner disc is reduced
significantly, coagulation should not play an important role in
the subsequent evolution of these grains.       

Figure\ref{SEDgaps} illustrates how the filtration process discussed above may influence
the disc SEDs.  The SED models in Figure \ref{SEDgaps} are computed assuming 
a 300 au flared disc of total mass $0.01 M_\odot$ that is passively heated 
by a star with $T_{\rm star} = 4000$ K, and $R_{\rm star} =
R_\odot$.  The disc surface density and scaleheight have radial
profiles of $\Sigma \propto r^{-1}$ and $h \propto r^{1.25}$, and the
scale height is normalised such that $h = 18$ au at $r = 100$ au. The temperature
is determined self-consistently by the Monte-Carlo radiation transfer calculation.
The solid curve is the SED resulting from a dust model with a grain size distribution of
$n(r) \propto r^{-3}$ with minimum and maximum grain sizes of $0.005 \micron$ and $1$ mm
respectively (see \citet{wood02}, Table 1 and Figure 3).  The other two models
simulate the dust filtration discussed above.  These models have the large grain model in
the outer disc (beyond 5 au) and ISM-like grains (maximum size $1 \micron$) inside 5 au.  The
dashed curve has a depletion factor of $10^4$ for the dust density, while the dotted
curve has a depletion factor of $10^6$, still within the possibilties discussed above.  
Notice the near-IR deficit and prominent silicate features at $10 \micron$ and $20 \micron$ for
the SED models with ISM-like grains and inner disc depletions.  The redistribution of IR excess
from short to long wavelengths for discs with optically thin interiors is also apparent as
discussed in \citet{rice03}. These SEDs are not models for any particular system, but demonstrate
the observable signatures of the filtration process we have described.  In these models, the
dust density in the inner disc is significantly reduced by filtration at the gap edge and by
accretion onto the planet, while the gas is only influenced by accretion onto the planet.
The gas density in the inner disc is therefore reduced by a much smaller amount than the dust density.
This extremely small dust-to-gas ratio in the inner disc produces an ``opacity gap", while still
allowing gas accretion onto the star, in line with observations of many T Tauri stars that exhibit
near-IR deficits in their SEDs.

It has, however, also been suggested \citep{eisner06} that ``opacity gaps" and 
the presence of small grains in the inner disc
could actually be evidence for grain growth and subsequent rapid inward
migration, rather than inferring the presence of a gap opening planet.  This conclusion
was partly because Jupiter-like planets in relatively massive discs should migrate into the central 
star within the disc lifetime \citep{eisner06}. Although this may indeed be the case, 
the viscous timescale at $5$ au can be $\sim 10^6$ years for reasonable viscosity values.  The observation
of such a system is therefore still possible, even if the planet ultimately doesn't survive the migration process. 
If near-IR deficits are due to grain growth, there must also be some
means of sharply truncating the grain 
size distribution at $\sim 1 \micron$ in the inner disc. It is
possible that we are seeing a population of planets that form 
early, and potentially do not actually survive the migration process.  

\section{Conclusion}
Although we cannot definitively conclude that near-IR deficits in the SEDs of some
TTauri stars are a consequence of an embedded Jupiter-like planet, that the filtering
effect of such a planet - producing a small grain population in the inner disc that
differs from the population in the outer disc - is consistent with SED modelling
of these systems is suggestive. The SED modelling also requires that the inner
disc be optically thin, while ongoing gas accretion requires that the
inner disc still have a substantial amount of gas.  If the grain size
distribution is somewhat flatter than the standard ISM size
distribution \citep{mathis77}, filtration at the outer gap edge can
significantly reduce the amount of solid material reaching the inner disc,
while still allowing sufficient gas through the gap. Although we don't actually know the grain size
distribution in the outer discs of those systems that show inner
holes, they do show evidence for grain growth. In some cases flatter size distributions do provide better fits
to TTauri SEDs \citep{dalessio01}, although this has yet to be tested
on the systems we consider here. That filtration of solid particles at
the outer egde of a gap opened by an embedded Jupiter-like planet can
produce not only a population of small grains in the inner disc, but
can potentially also make this region optically thin, makes this an
attractive possibility.

The filtration process should also lead to an enhancement of 
larger particles at the peak of the gap edge. Particles exterior to the gap will continue
to drift inwards until they reach the gap edge, at which point the change in the gas
pressure gradient will prevent the larger particles from drifting any further. If the 
surface density of such particles becomes sufficiently high, further planet formation could
occur at this location.  
Since `Hot' Jupiters are thought to form at modest radii and then migrate inwards via Type II migration
to their present locations, one may expect in some cases to find additional lower mass 
planets in 2:1 resonances with these `Hot' Jupiters.  The presence of even terrestrial mass planets in 
resonant orbits with `Hot' Jupiters could be detected using transit timing \citep{agol05}. Although
such a detection would be consistent with the analysis presented here, it has been suggested that
this could also occur if low-mass, fast migrating planets 
($> 0.1$ M$_\oplus$) were resonantly trapped by gap opening planets \citep{thommes05}.  In the latter case, however,
one may expect the trapped planets to subsequently grow into gas giants \citep{thommes05}, 
while in the former case the planetesimal trapping could occur later when there may only be sufficient time
for the newly forming planet to reach a terrestrial-like mass.

\section*{acknowledgements}
The authors would like to acknowledge interesting discussions with
Richard Alexander, and would like to thank the referee,
C.P. Dullemond, for some extremely useful and insightful suggestions. 
This work was supported by NASA under grants NAG5-13207 and NNG04GL01G 
from the Origins of Solar Systems and Astrophysics Theory Programs, and 
by the NSF under grant AST~0407040.

\end{document}